\documentclass[useAMS,usenatbib]{mn2e}
\usepackage{graphicx,natbib}
\usepackage{psfrag}
\usepackage{amssymb}
\usepackage{float}
\usepackage[dvips]{color}
\usepackage[export]{adjustbox}  
\usepackage{multicol}
\usepackage{supertabular}
\usepackage{changepage}
\usepackage{pdflscape}
\usepackage{lmodern}
\usepackage{lscape}
\usepackage{longtable}
\usepackage{dcolumn}

\title[Three GCs discovered in the bulge]{Three candidate globular clusters discovered in the Galactic bulge}

\author[D. Camargo and D. Minniti]{D. Camargo$^{1}$ and  D. Minniti$^{2,3,4}$\\
$^1$ Col{\'e}gio Militar de Porto Alegre, Minist{\'e}rio da Defesa, 
Av. Jos{\'e} Bonif{\'a}cio 363,
Porto Alegre 90040-130, RS, Brazil\\
$^2$ Departamento de Fisica, Facultad de Ciencias Exactas, Universidad Andres Bello, Av. Fernandez Concha 700, Las Condes,\\
Santiago, Chile\\
$^3$ Instituto Milenio de Astrof{\'i}sica, Santiago, Chile\\
$^4$ Vatican Observatory, V00120 Vatican City State, Italy}

\begin{document}

\pagerange{\pageref{firstpage}--\pageref{lastpage}}

\maketitle

\label{firstpage}

\begin{abstract}
This work reports the discovery of three new globular clusters (GCs) towards the Galactic bulge - Camargo 1107, 1108, and 1109. The discovery was made using the WISE, 2MASS, VVV, and Gaia-DR2 photometry.  
The new findings are old ($12.0-13.5$ Gyr) and metal-poor GCs ($[Fe/H]<-1.5$ dex) located in the bulge area close to the Milky Way (MW) mid-plane. 
Although the old ages and low metallicities suggest that the newly discovered GCs are likely associated with the inner halo the possibility of these clusters being part of a primordial bulge GC subpopulation cannot be ruled out. Camargo 1107, for instance, presents a metallicity of $[Fe/H]=-2.2\pm0.4$ dex and an age of $13.5\pm2$ Gyr, which may suggest that this cluster formed just after the Big Bang in the very early Universe. The discovery of GCs such as the new findings is crucial to built a coherent picture of the inner Galaxy. It is likely that at least a few more dozens of GCs are still to be discovered in the bulge.
\end{abstract}


\section{Introduction} \label{sec:intro}

Recently \citet{Camargo18} communicated the discovery of five new globular clusters (GCs) in the Milky Way bulge. The VVV team also listed almost a hundred of new globular cluster candidates in the last few years \citep{Minniti11, Moni11, Minniti17a, Minniti17b, Minniti17c, Borissova18, Barba18}, some of them analyzed by \citet{Piatti18}. These newly discovered GCs complement the \citet{Harris96} compilation that contain 157 entries. \citet{Ryu18} found two new GCs and provided a list with the recent discoveries.

As relics of star formation in the early Universe, globular clusters (GCs) may provide important clues on the Milky Way history. For instance, the bulge formation and evolution remains poorly understood and GCs are powerfull tools to trace its structure, kinematics, and stellar content. 
The bulge has been the subject of an active debate in the last few years, which generated an important effort in order to  characterize properly the central region of our home Galaxy \citep{Valenti07, Bica16, Cohen17, Barbuy18, Kerber18, Rossi18}. The near- and mid-IR photometry provided by  wide-field sky surveys\footnote{such as 2MASS, WISE, DECaPS, ESO-VVV, Gaia, and more recently ESO-VVVX and Gaia-DR2} is boosting our knowledge on this issue, since the bulge direction is heavily obscured by dust and stellar crowding in the visible wavelengths. In this way, accurate ages and metallicities are need to reconstruct the bulge's history from its formation to the present-day.

In the current view bulges are classified in \textit{classical bulges} and \textit{disk-like bulges} \citep{Kormendy04}. Classical bulges are thought to emerge from violent events such as galaxy mergers or sinking of giant gas clumps and host older stellar population within a spherical structure resembling elliptical galaxies. Flattened disk-like bulges may arise on longer timescales via  internal processes such as disk instabilities and secular evolution.

This study announces the discovery and measures the parameters of three new GCs projected towards the Galactic bulge, adding entries to the star cluster catalog built by a survey on the WISE stellar maps \citep{Camargo15a, Camargo15b, Camargo15c, Camargo16a, Camargo16b}. Thus, the newly discovered GCs are named Camargo 1107, Camargo 1108, and Camargo 1109.

\begin{figure*}
\centering
\begin{minipage}[b]{1.0\linewidth}
\begin{minipage}[b]{1.0\linewidth}
\includegraphics[width=0.36\linewidth]{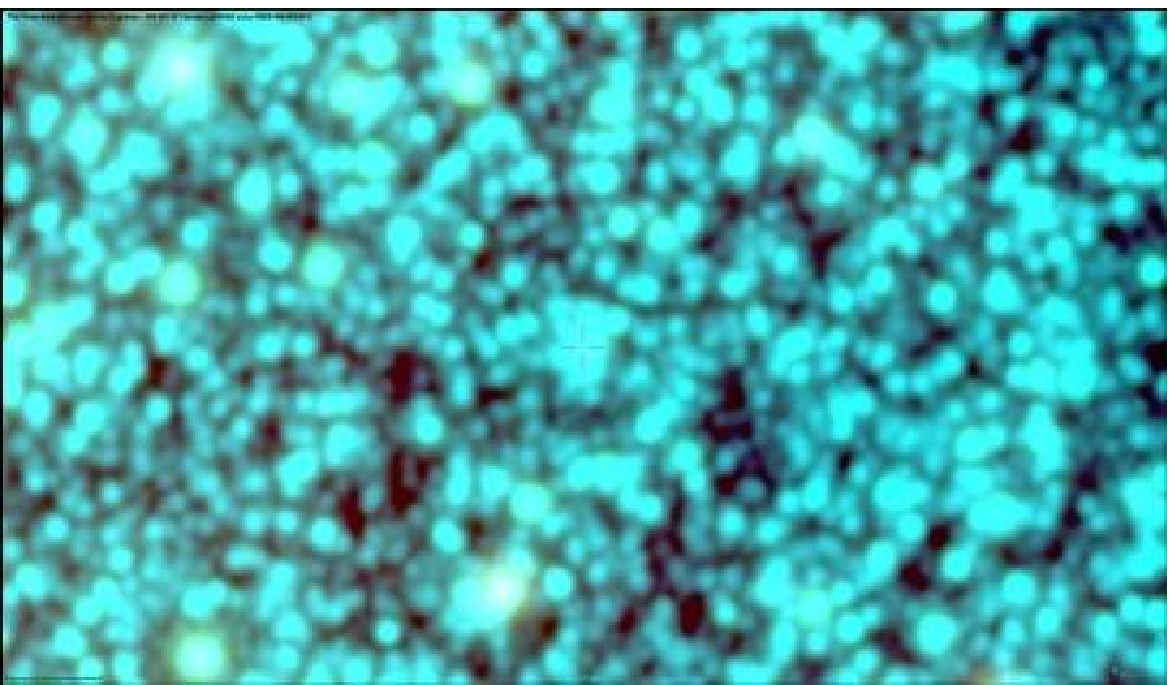}
\includegraphics[width=0.32\linewidth]{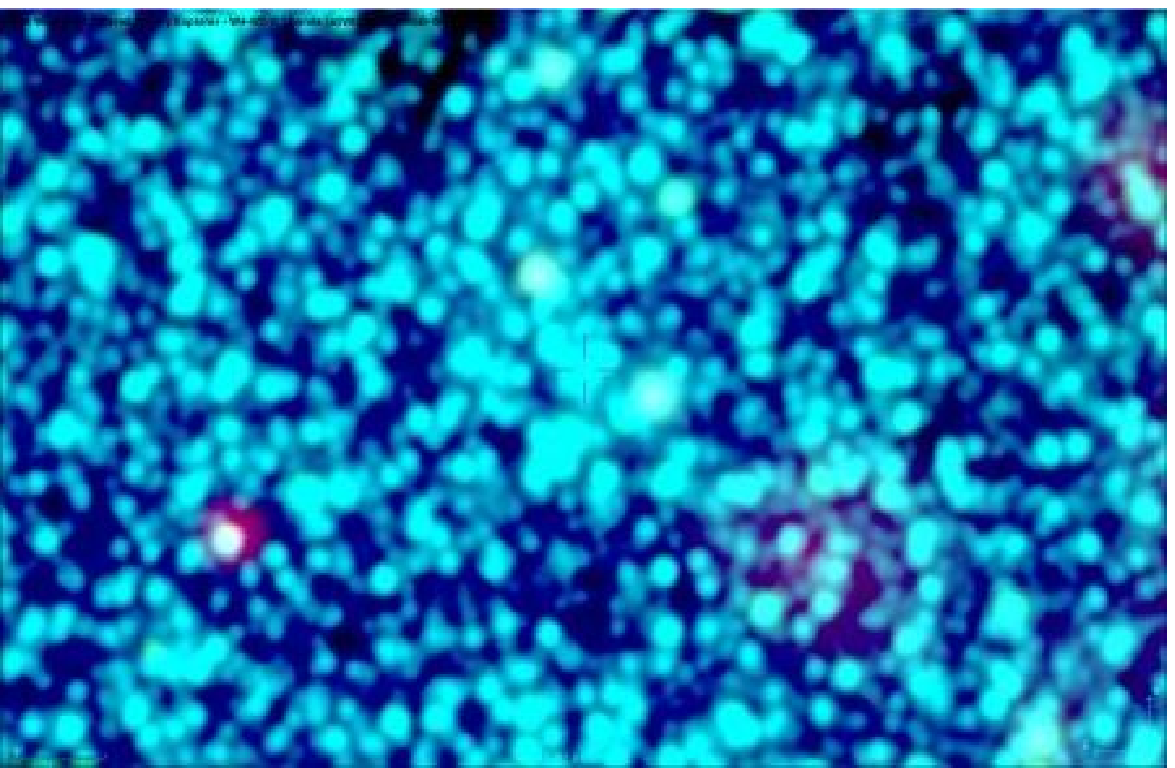}
\includegraphics[width=0.32\linewidth]{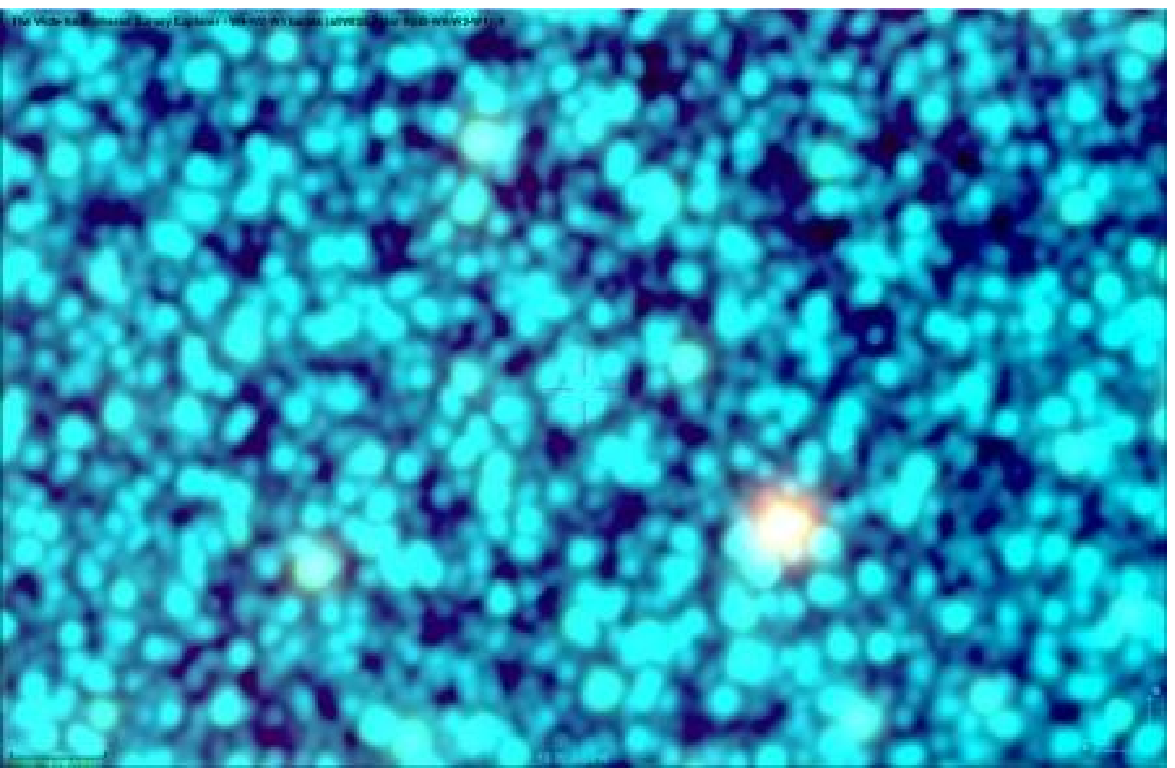}
\end{minipage}\hfill
\end{minipage}\hfill
\caption{WISE multicolor images of the central regions $(12.3'\,\times\,8.0')$ of the new GCs. From right to left Camargo 1107, Camargo 1108, and Camargo 1109. North
is to the top and east to the left.}
\label{f1}
\end{figure*}

\section{Methods}
\label{sec:2}

The VVV and 2MASS photometry in the $J$, $H$ and $K_s$ bands are used to analyse the nature of the newly discovered GCs.

The 2MASS color-magnitude diagrams (CMDs) are built by using a field-star decontamination procedure \citep[][]{Bonatto07, Camargo15a, Camargo15b, Camargo16a}. The analysis also employs the Gaia-DR2 photometry \citep{Gaia18} to build the proper motion diagrams (PMDs) used to reinforce the cluster nature of the new objects.

The basic parameters are derived via PARSEC isochrones fitting \citep{Bressan12} to the field-star decontaminated CMDs. Such a procedure is guided by the  direct comparison with the CMD of a selected cluster \citep{Camargo18}.

\begin{figure*}
\centering
\begin{minipage}[b]{0.8\linewidth}
\includegraphics[width=\textwidth]{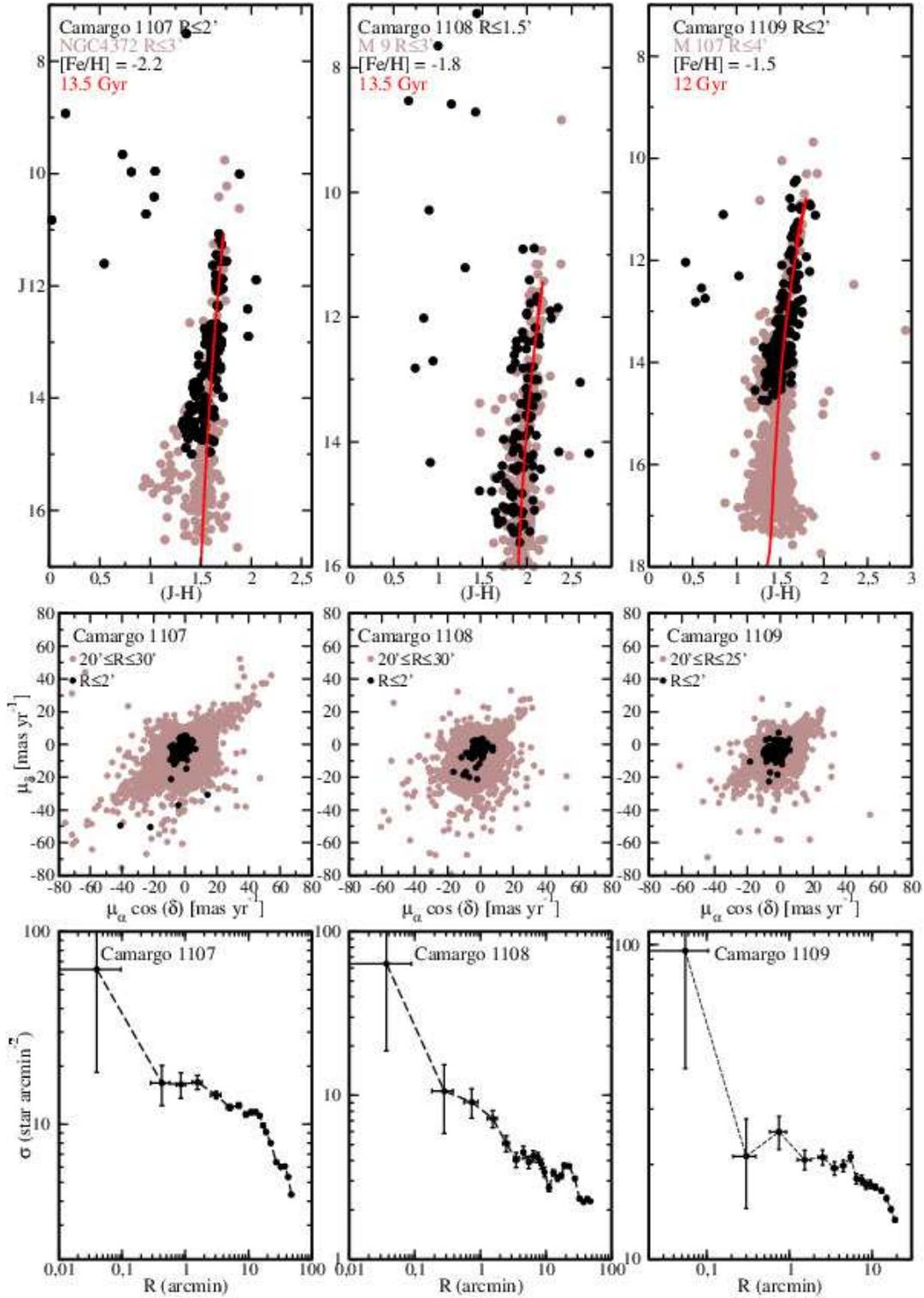}
\end{minipage}\hfill
\caption{Top panels: 2MASS field-star decontaminated CMDs of the newly discovered GCs and their respective reference clusters.  The best-fitting PARSEC isochrones
are shown as solid lines. Middle panels: Gaia-DR2 PM distribution. The black circles are the stars in the central region of each cluster while the brown circles represent the stars in the surrounding field. Bottom panels: 2MASS RDPs built after applying color-magnitude filters that select the RGB stars fitted by the isochrones.}
\label{f2}
\end{figure*}

The 2MASS radial density profiles (RDPs) of the new findings (Fig.~\ref{f2}, bottom panels) are built by applying color-magnitude filters to the observed photometry.

\begin{figure}
\centering
\begin{minipage}[b]{0.8\linewidth}
\includegraphics[width=\textwidth]{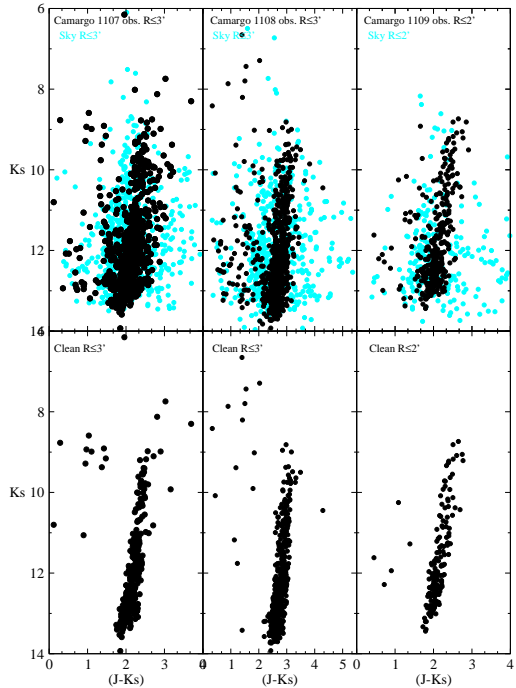}
\end{minipage}\hfill
\caption{Top panels: 2MASS observed CMDs extracted from the central regions of the newly discovered GCs and the respective  equal-area comparison field.  Bottom panels: field-star decontaminated
CMDs.}
\label{f3}
\end{figure}

\section{Present discoveries}
\label{sec:3}

\begin{figure*}
\centering
\begin{minipage}[b]{1.0\linewidth}
\begin{minipage}[b]{1.0\linewidth}
\includegraphics[width=0.329\linewidth]{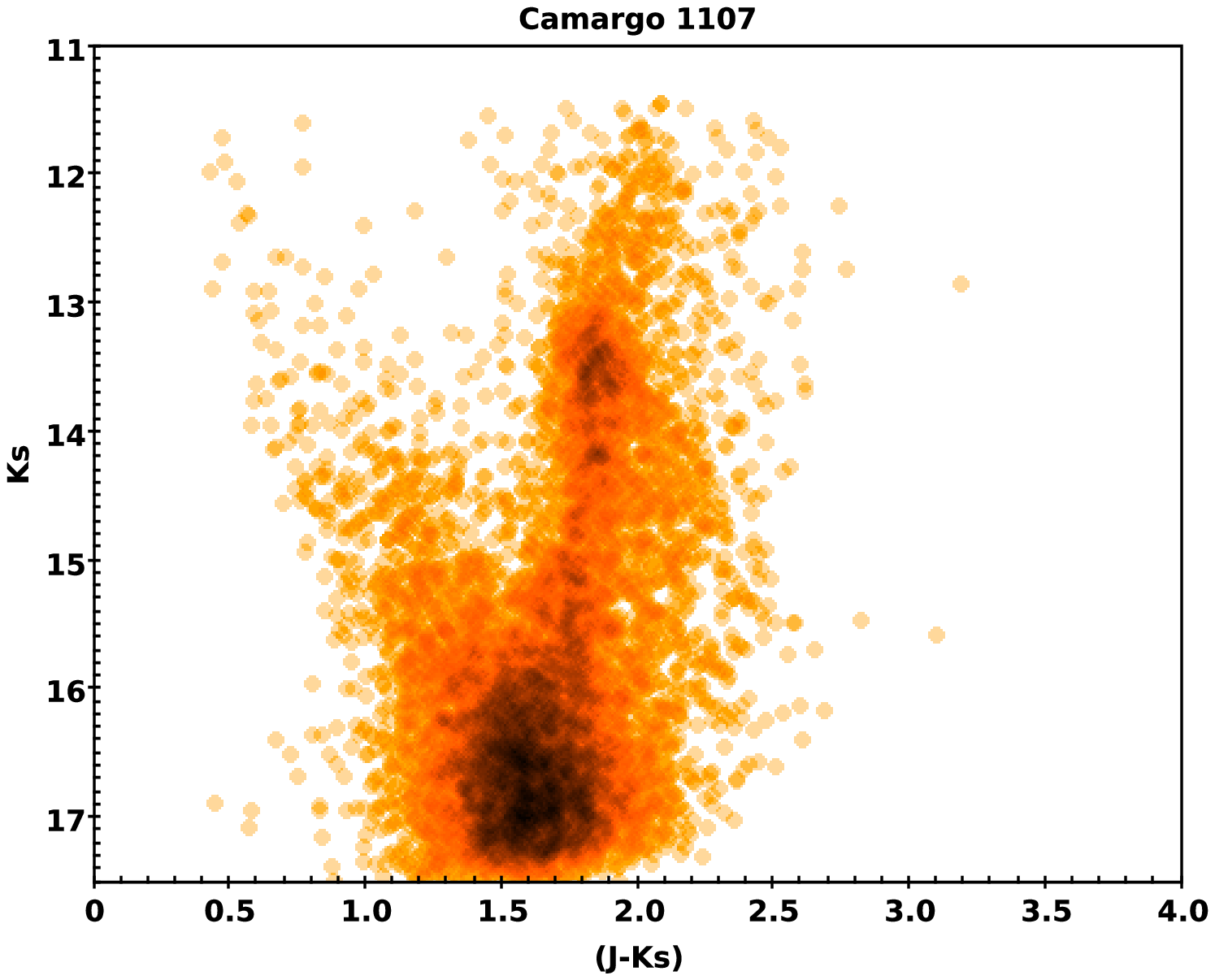}
\includegraphics[width=0.33\linewidth]{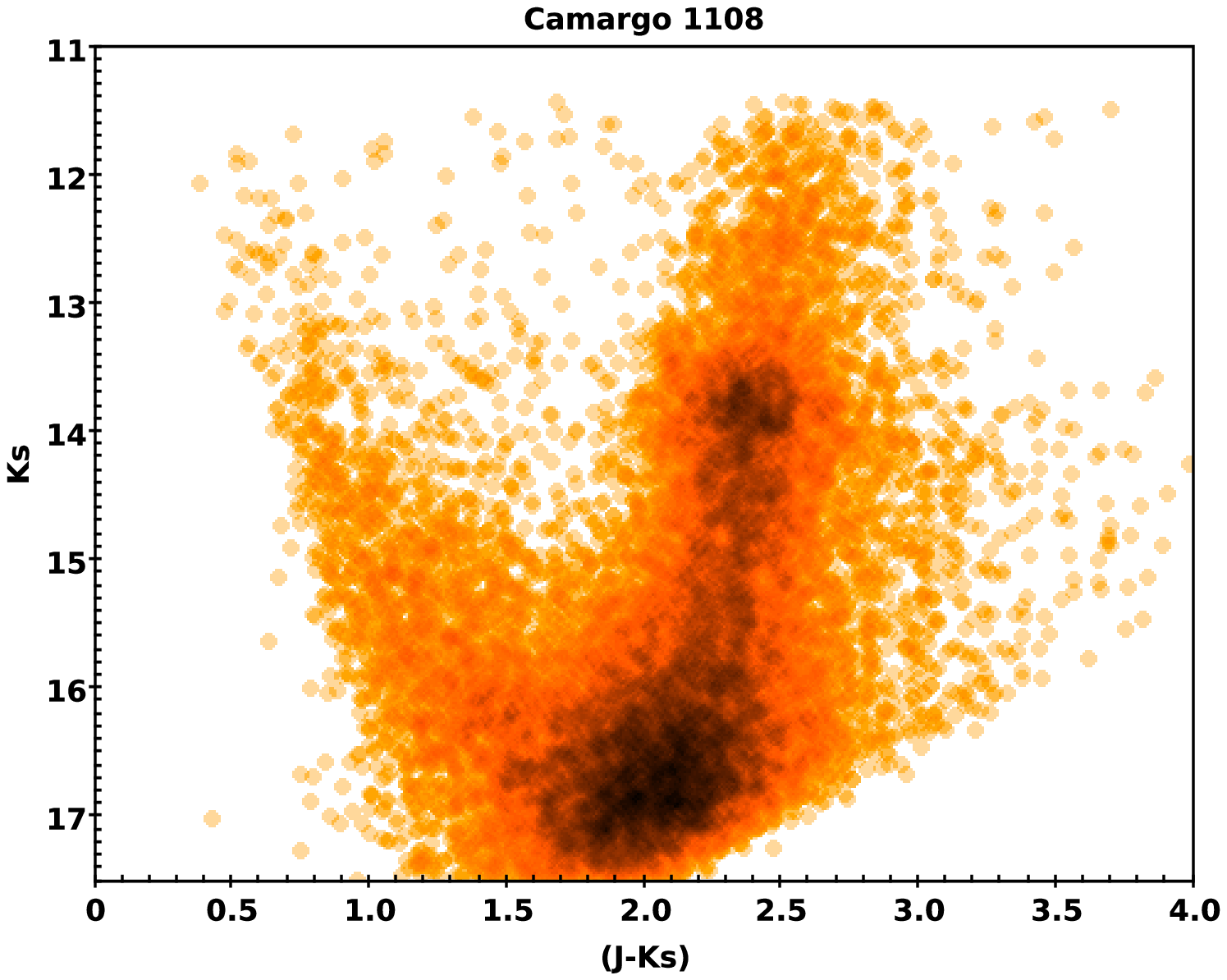}
\includegraphics[width=0.33\linewidth]{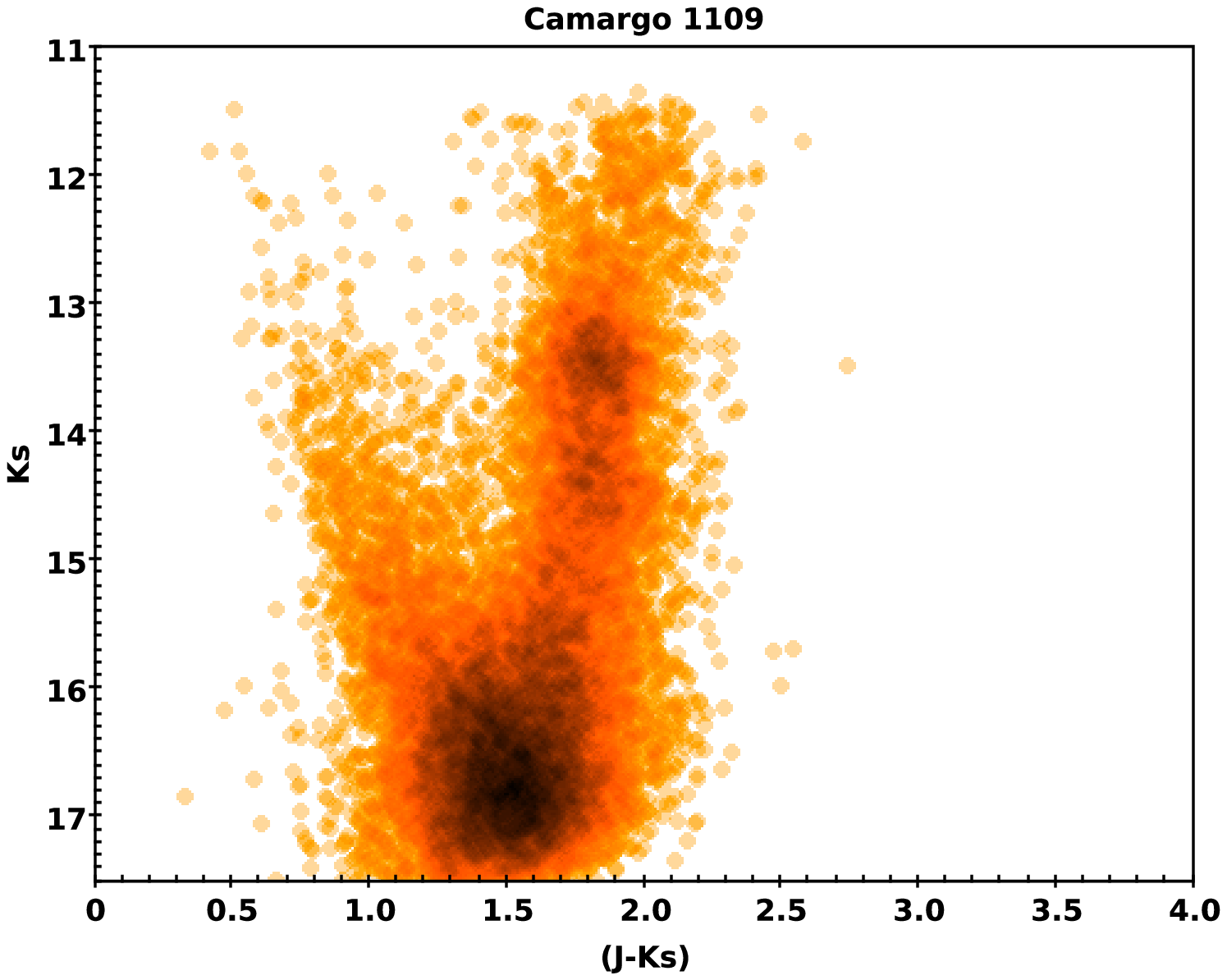}
\end{minipage}\hfill
\end{minipage}\hfill
\caption{The VVV $K_s\times(J-K_s)$ observed CMDs ($R\leq1.5'$) of the new globular cluster candidates Camargo 1107, Camargo 1108, and Camargo 1109.}
\label{f4}
\end{figure*}

The WISE multicolor images for the new findings are shown in Fig.~\ref{f1} whereas Figs.~\ref{f3}, ~\ref{f2}, and ~\ref{f4} provide the CMDs for probable cluster members, which are basically populated by RGB stars, given the crowding towards the bulge \citep{Camargo18}. The bluer and brighter stars in the decontaminated CMDs of the new findings are foreground stars that survived a non-$100\%$ efficient subtraction. The statistical field-star decontamination is affected by the crowding and high extinction towards the bulge. The middle panels of the Fig.~\ref{f2} show the Gaia-DR2 PM for stars located within the GC central region (black circles), after applying a filter that discard stars with PM uncertainties $\geq\,0.5$ $mas\, yr^{-1}$. These panels also show the stars within the respective comparison fields (brown circles). In the bottom panels of Fig.~\ref{f2} are shown the 2MASS RDPs built after applying color-magnitude filters that select the RGB stars fitted by the isochrones in the magnitude range from the bottom of the distribution of the decontaminated stars to the top of the respective isochrone. The RDPs of the new findings reinforce their GC nature, but deeper photometry is required to derive accurate parameters.

\begin{table*}
\begin{center}
\caption{Galactic coordinates and basic parameters for the GCs discovered in the present study.}
\renewcommand{\tabcolsep}{2.9mm}
\renewcommand{\arraystretch}{1.3}
\begin{tabular}{lrrrrrrrrr}
\hline
\hline
Cluster&$\ell$&$b$&$A_V$&$M_V$&Age&$[Fe/H]$&$d_{\odot}$&$R_{GC}$&Size\\
&$(^{\circ})$&$(^{\circ})$&(mag)&(mag)&(Myr)&dex&(kpc)&(kpc)&$^{\prime}$\\
\hline
Camargo 1107 &$357.977$&$0.956$&$3.3\pm0.1$&$-6.6\pm0.5$&$13.5\pm2.0$&$-2.2\pm0.4$&$4.0\pm0.7$&$4.29\pm0.74$&10\\
Camargo 1108 &$358.404$&$-1.087$&$4.5\pm0.03$&$-8.4\pm0.5$&$13.5\pm1.5$&$-1.8\pm0.3$&$3.3\pm0.5$&$5.0\pm0.47$&8\\
Camargo 1109 &$2.165$&$0.844$&$3.3\pm0.1$&$-6.4\pm0.7$&$12.0\pm1.5$&$-1.5\pm0.2$&$4.3\pm0.6$&$3.97\pm0.61$&7\\
\hline
\end{tabular}
\begin{list}{Table Notes.}
\item The $R_{GC}$ is calculated adopting a Heliocentric distance of $R_{\odot}=8.3$ kpc. 
\end{list}
\label{tab1}
\end{center}
\end{table*}

\subsection{Camargo 1107}

The Galactic coordinates of the newly discovered GC Camargo 1107 are $\ell=357.977^{\circ}$ and $b=0.956^{\circ}$.
The WISE multicolor image centred in the coordinates of the new GC (Fig.~\ref{f1}) presents a relatively clear stellar concentration in the cluster central region.
The decontaminated CMDs of this GC are shown in Figs.~\ref{f3} and ~\ref{f2}. In addition to the best isochrone solution the CMD of the newly discovered GC is compared to the NGC 4372 decontaminated CMD (Fig.~\ref{f2}). NGC 4372 presents a metallicity of $[Fe/H]=-2.17$  for an age of $15\pm4$ Gyr \citep{Alcaino91, Harris96}. The analysis provides an extinction of $A_V=3.3\pm0.3$ mag and suggest that Camargo 1107 is an old and very metal-poor GC with a metallicity of $[Fe/H]=-2.2\pm0.4$ dex and an age of $13.5\pm2.0$ Gyr.  The distance from the Sun is $d_{\odot}=4.0\pm0.7$ kpc and the Galactocentric distance  $R_{GC}=4.29\pm0.74$ kpc, adopting a Heliocentric distance of $R_{\odot}=8.3$ kpc. The rectangular coordinates derived are $x_{GC}=-4.29\pm0.74$ kpc, $y_{GC}=-0.14\pm0.03$ kpc, and a vertical distance from the Galactic plane of $z_{GC}=0.067\pm0.01$ kpc. This newly discovered GC presents an absolute magnitude of $M_V=-6.6\pm0.5$ mag. The old age and extremely low metallicity of Camargo 1107 points to an inner halo GC. The Gaia PM (Fig.~\ref{f2}) for the stars in the central region of Camargo 1107 follow the distribution expected for a bulge/inner halo population \citep{Bobylev17}. As for the other GCs discovered in this study, the bright stars in the CMD seem to be field stars that remained after the decontamination procedure. The VVV CMD (Fig.~\ref{f4}) reinforces the cluster nature of Camargo 1107.

\subsection{Camargo 1108}

The decontaminated CMDs of Camargo 1108 are shown in the Figs.~\ref{f3} and \ref{f2}. The best fitting isochrone and direct comparison with the M 19 RGB (Fig.~\ref{f2}) provides the basic parameters. M 19 is an inner halo globular cluster with $[Fe/H]\sim-1.74$ \citep{Harris96}.
Camargo 1108 with Galactic coordinates $\ell=358.404^{\circ}$ and $b=-1.087^{\circ}$  is a $13.5\pm1.5$ Gyr GC located at a distance from the Sun of $d_{\odot}=3.3\pm0.5$ kpc and a Galactocentric distance  $R_{GC}=5.0\pm0.5$ kpc. The rectangular coordinates derived are $x_{GC}=-5.0\pm0.47$ kpc, $y_{GC}=-0.09\pm0.01$ kpc, and $z_{GC}=-0.063\pm0.009$ kpc. This GC presents an absolute magnitude of $M_V=-8.4\pm0.5$ mag. The extinction in the cluster direction is $A_V=4.5\pm0.03$ mag. The metallicity of Camargo 1108 is $[Fe/H]=-1.8\pm0.3$ dex.

\subsection{Camargo 1109}

The Galactic coordinates of Camargo 1109 are $\ell=2.165^{\circ}$ and $b=0.844^{\circ}$.
The nature of this GC is established by using M 107 as a reference cluster. M 107 presents an age ranging from $12-12.75$ Gyr and $[Fe/H]\sim-1.0$ \citep{Dotter10, VandenBerg13, Harris96}.
The comparison of the decontaminated CMDs of Camargo 1109 and M 107 plus the isochrone solution (Fig.~\ref{f2}) point to an old and metal-poor GC with age of $12\pm1.5$ Gyr and  $[Fe/H]=-1.5\pm0.2$ dex. The extinction for this GC is $A_V=3.3\pm0.1$ mag. The absolute magnitude of Camargo 1109 is $M_V=-6.4\pm0.7$ mag.
Located at a distance from the Sun of $d_{\odot}=4.3\pm0.6$ kpc Camargo 1109 presents a Galactocentric distance of  $R_{GC}=3.97\pm0.61$ kpc and rectangular coordinates of $x_{GC}=-3.96\pm0.61$ kpc, $y_{GC}=0.16\pm0.02$ kpc, and $z_{GC}=0.064\pm0.009$ kpc.

\section{SUMMARY AND DISCUSSION}
\label{sec:5}

Since globular clusters basically witnessed the entire history of our Galaxy, they may allow us to reconstitute the chain of physical processes experienced by the Milky Way from its origin to the \textit{present-day}. However, the census of globular clusters in the Milky Way is still far from complete, especially for the bulge GCs.
In this context, recently \citet{Camargo18} communicated the discovery of five new GCs within the bulge (Camargo 1102 to 1106). 
This study reports the discovery of another three new GCs towards the bulge - Camargo 1107, Camargo 1108, and Camargo 1109. 

The newly discovered GCs are very old and extremely metal-poor, for their location, with ages in the range of $12.0-13.5$ Gyr and $[Fe/H]$ ranging from $-1.5$ to $-2.2$ dex. They are located within $3.9-5.8$ kpc from the MW center and are close to the Galactic mid-plane. Camargo 1107 is a special case, although the fit of younger and more metal-rich isochrones is possible the best fit suggests an age of $13.5$ Gyr and a metallicity of $[Fe/H]=-2.2$ dex (Table~\ref{tab1}). The GCs Camargo 1107 and Camargo 1108 have ages and metallicities similar to the reference GCs, but the best isochrone solution suggest that Camargo 1109 is more metal-poor than M 107.

The GCs discovered in this study, just as those in the previous paper \citep{Camargo18} suggest that the MW central region hosts a subpopulation of very old and metal-poor GCs, which is consistent with being an inner halo component \citep{Bica16, Kilic17, Nogueras18, Villegas17}.  
 Alternatively, these clusters may be part of an old classical bulge built up by merging in the early MW history \citep{Minniti16, Pietrukowicz15, Rossi18, Kruijssen18}.  In this sense, the Gaia-DR2 PM for member-stars of the new findings follow the bulge ones. It is expected that halo globular clusters present high-proper motion distribution relative to the Galactic disk/bulge clusters and the stellar background.
 
These clusters may be the remaining of a primordial class of GCs that  were destroyed mainly by dynamical processes and are the source of the ancient field stars that inhabit the Milky Way bulge and the inner halo \citep{Minniti16, Dong17, Barbuy18}.
Previous works suggest that there are a significant fraction of GCs which were formed just after the Big Bang, around the epoch of reionization \citep{Forbes18}. 

Given the potential of the newly discovered GCs as fossils of the primordial Universe deeper photometry reaching the MS TO is necessary for deriving accurate ages and metallicities. In addition, it seems that many faint globular clusters such as the present discoveries remain undetected until now due to the high extinction and stellar crowding towards the bulge.

\vspace{0.8cm}

\textit{Acknowledgements}: 
We thank an anonymous referee for useful remarks that improved the paper. 
We gratefully acknowledge data from the ESO Public Survey program ID $179.B-2002$ taken with the VISTA telescope, and products from the Cambridge Astronomical Survey Unit (CASU). DM is supported by the BASAL Center for Astrophysics and Associated Technologies (CATA) through grant $AFB-170002$, by the Ministry for the Economy, Development and Tourism, Programa Iniciativa Cient\'{i}fica Milenio grant $IC120009$, awarded to the Millennium Institute of Astrophysics (MAS), and by FONDECYT $No.\,1170121$.
 This publication also makes use of data products from the Wide-field Infrared Survey Explorer (WISE), Two Micron All Sky Survey (2MASS), and {\it Gaia}-DR2.

\end{document}